\documentclass[a4paper]{EslabStyle}
\usepackage{graphics}

\begin{document}

\title{Cosmic Star Formation History Required from 
Infrared Galaxy Number Count : 
Future Prospect for {\sl Infrared Imaging Surveyor} ({\sl IRIS})
}

\author{T.\,T. Ishii\inst{1,2} \and T.\,T. Takeuchi\inst{1} \and 
H. Hirashita\inst{1} \and K. Yoshikawa\inst{1}}

\institute{Department of Astronomy, Faculty of Science, Kyoto University, 
JAPAN, 606-8502
\and 
Kwasan and Hida Observatories, Kyoto University, JAPAN, 607-8471
}

\maketitle

\begin{abstract}
We constructed a model of infrared and sub-mm (hereafter IR) 
galaxy number count and estimated 
history of the IR luminosity density. 
We treat the evolutionary change of galaxy luminosities 
as a stepwise nonparametric form, in order to explore the most suitable
evolutionary history which reproduces the present observational results.
We found the evolutionary patterns which satisfy 
both constraints required from Cosmic Infrared Background (CIRB) 
and IR galaxy number counts.
One order of magnitude increase of luminosity 
at redshift $z=0.75 - 1.0$ was found in IR $60\; \mu$m 
luminosity density evolution.
We also found that a large number of galaxies 
( $\sim 10^7$ in the whole sky) will be 
detected in all-sky survey at far-infrared by 
{\sl Infrared Imaging Surveyor} ({\sl IRIS}); 
Japanese infrared satellite project Astro-F.
\end{abstract}

\section{Introduction}

Recent infrared and sub-mm surveys revealed a very steep
slope of galaxy number count compared with that expected
from no evolution model, and provided a new 
impetus to the related field.
Such excess of galaxy number count is generally understood 
as a consequence of strong galaxy evolution, i.e. rapid 
change of the star formation rate in galaxies.
Now 
Japanese infrared satellite project Astro-F 
({\sl Infrared Imaging Surveyor}: {\sl IRIS}) is in 
progress, and we calculated the expected number count by
a simple empirical method (Takeuchi et al. 1999; Hirashita 
et al. 1999).
The applied model was based on the {\sl IRAS} surveys, 
and we need more realistic one to study the detailed 
observational plans and follow-up strategies.
In order to construct the advanced model, we first compiled the 
infrared/sub-mm SEDs of galaxies 
obtained by {\sl ISO}, SCUBA, and other facilities, and 
derived average SEDs for various classes of galaxies.
Then, 
using the local luminosity function, we studied the 
required galaxy evolutionary history statistically.

\begin{figure}[ht]
\resizebox{\hsize}{!}{\includegraphics{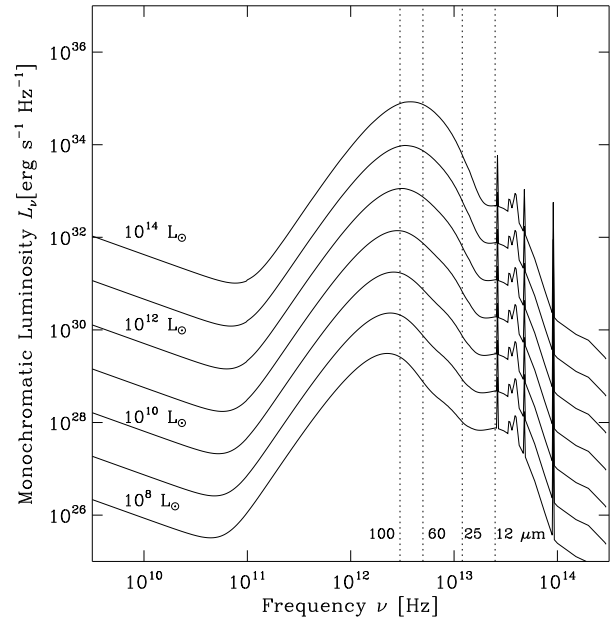}}
\caption{Assumed galaxy spectral energy distribution (SED) 
in the near infrared to radio wavelengths. \label{fig1}}
\end{figure}

\begin{figure}[ht]
\resizebox{\hsize}{!}{\includegraphics{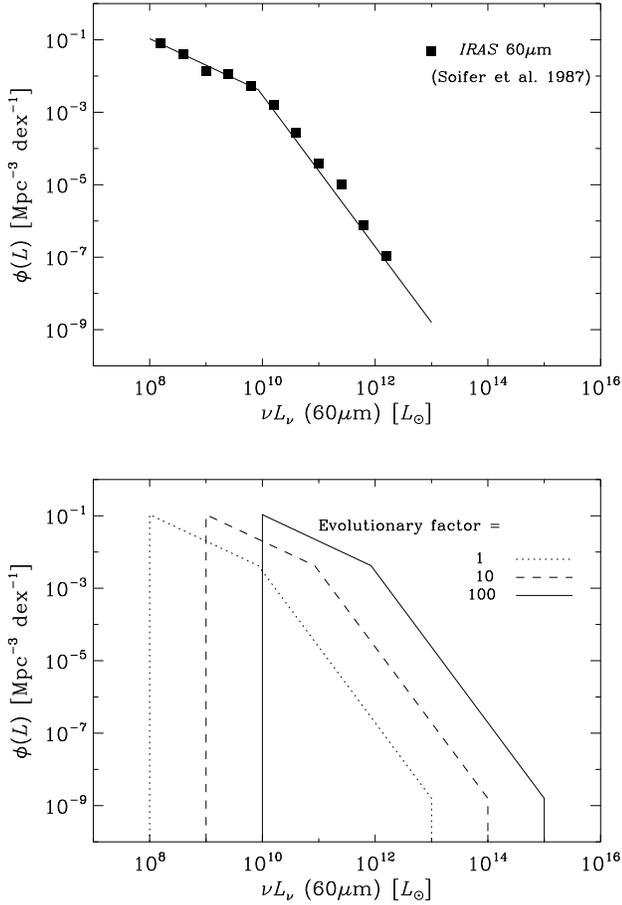}}
\caption{The double power-law form (Soifer et al. 1987) for the local 
luminosity function. 
We assumed pure luminosity evolution in this study. \label{fig2}}
\end{figure}

\begin{figure}[ht]
\resizebox{\hsize}{!}{\includegraphics{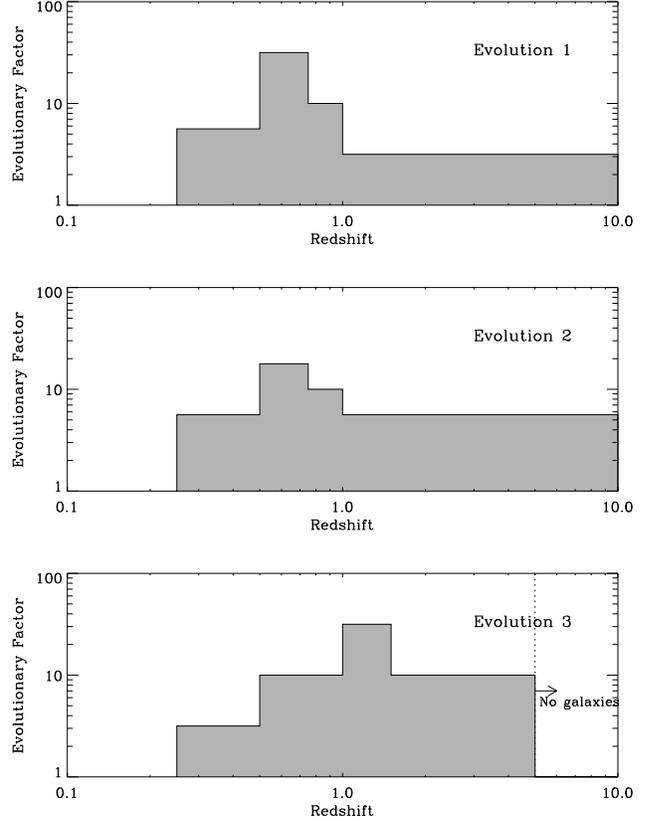}}
\caption{Stepwise nonparametric form of 
the evolutionary change of galaxy luminosities.
 \label{fig3}}
\end{figure}

\begin{figure}[ht]
\resizebox{\hsize}{!}{\includegraphics{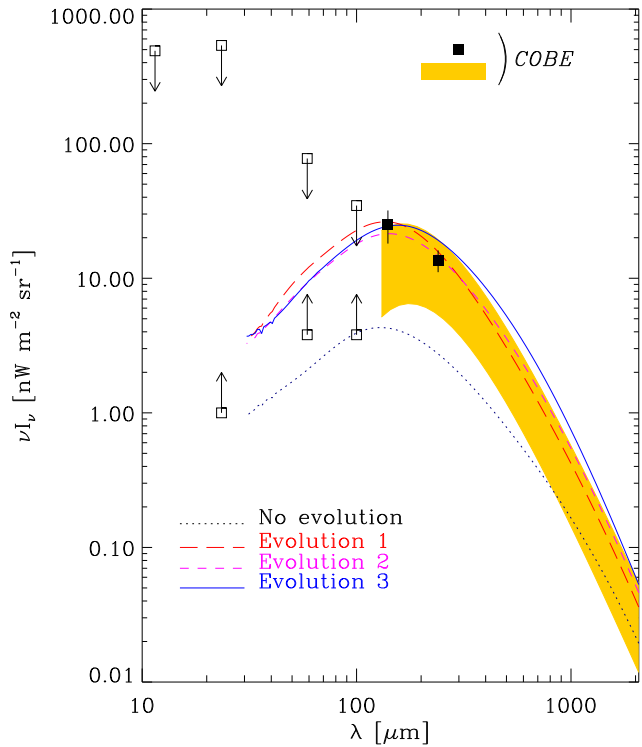}}
\caption{The expected cosmic infrared background (CIRB)
spectra with Evolution $1-3$ and No evolution.
The observational constraints on CIRB obtained by 
COBE measurement (Fixsen et al. 1998, 
Hauser et al. 1998 and references therein) 
are also plotted. \label{fig4}}
\end{figure}

\begin{figure}[ht]
\resizebox{\hsize}{!}{\includegraphics{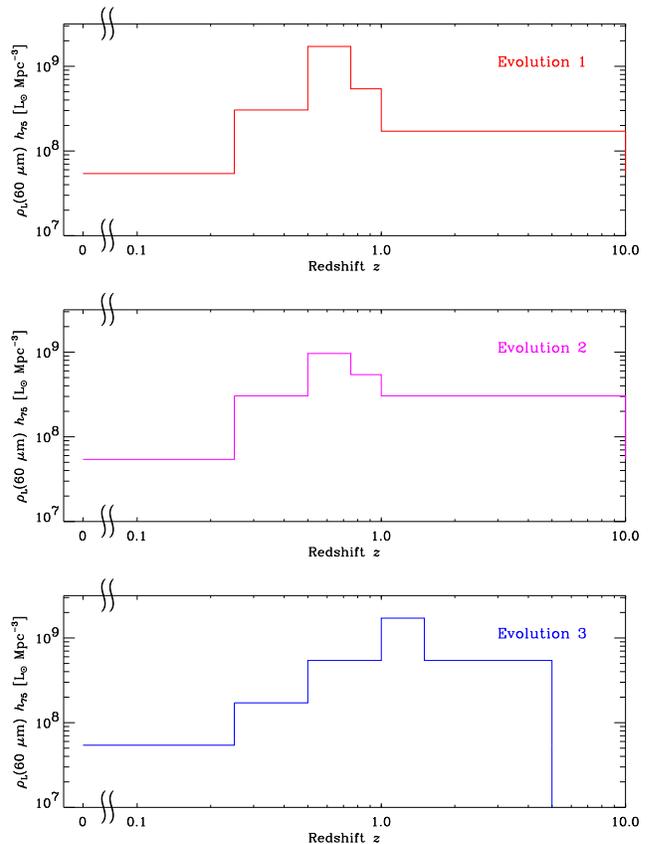}}
\caption{{\it The FIR $60\; \mu$m luminosity density evolution.} \label{fig5}}
\end{figure}

\begin{figure}[ht]
\resizebox{\hsize}{!}{\includegraphics{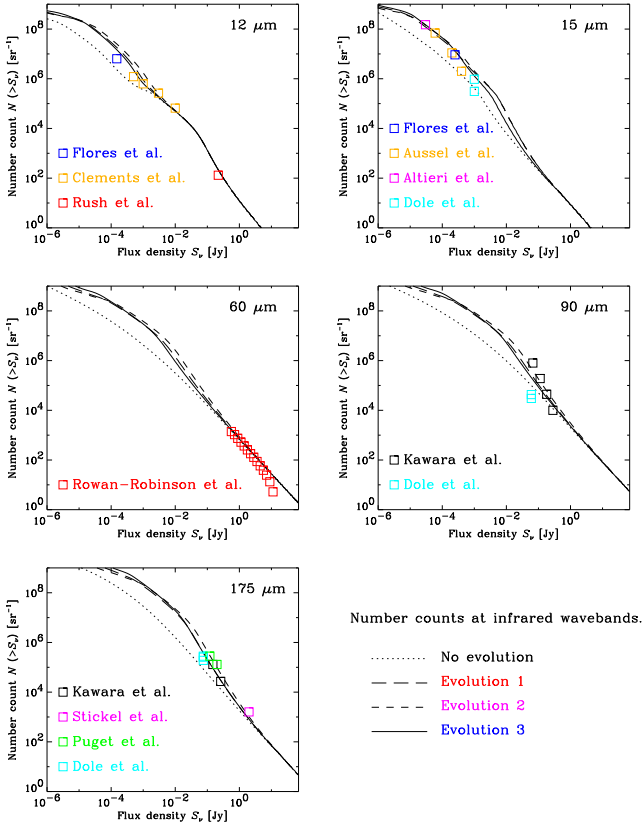}}
\caption{{\it The multiband galaxy at the infrared
wavelengths.} \label{fig6}}
\end{figure}

\begin{figure}[ht]
\resizebox{\hsize}{!}{\includegraphics{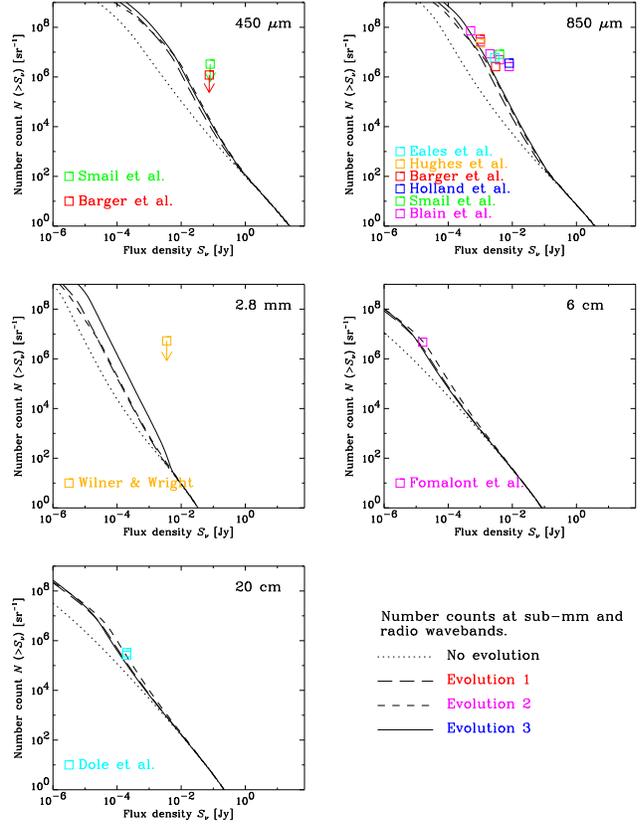}}
\caption{The multiband galaxy at the sub-mm -- radio
wavelengths. \label{fig7}}
\end{figure}

\begin{figure}[ht]
\resizebox{\hsize}{!}{\includegraphics{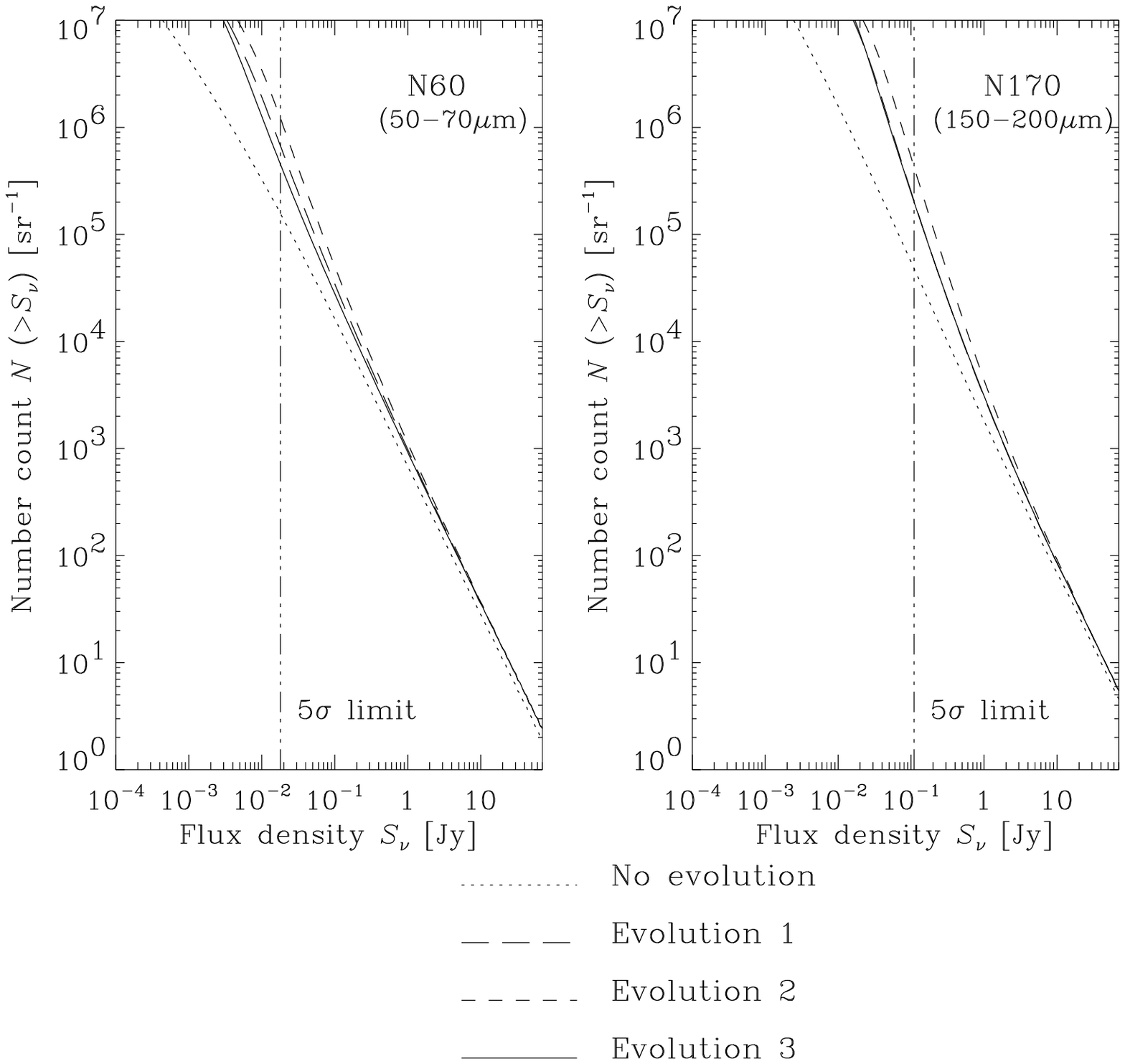}}
\caption{The galaxy number count predictions at assumed 
IRIS two bandpasses (N60 and N170).
The dot-dashed lines denote the IRIS  
Far-infrared Scanner (FIS) flux detection 
limit at each waveband.
 \label{fig8}}
\end{figure}

\section{Model Description}

We construct the SEDs of galaxies based on the {\sl IRAS}~color--luminosity 
relation (Smith et al. 1987; Soifer \& Neugebauer 1991).
For the infrared -- sub-mm component, we consider PAH (polycyclic aromatic 
hydrocarbon), graphite and silicate dust spectra (Dwek et al. 1996).
Detailed PAH band emission parameters are taken from Allamandola et al. 
(1989).
For the longer wavelength regime, power-law continuum produced by 
synchrotron radiation ($\propto \nu^{-\alpha}$) dominates.
We set $\alpha = 0.7$ according to Condon (1992).
The SEDs are presented in Fig. 1.

We applied the double power-law form (Soifer et al. 1987) for the local 
luminosity function, and assumed pure luminosity evolution in this study.
This is depicted in Fig. 2.
The faint-end slope does not affect the result, because the number 
count is an integrated value along with redshift and volume, and its
contribution is small.

\section{Results and Discussions}

\subsection{Evolutionary history}

We treat the evolutionary change of galaxy luminosities 
as a stepwise nonparametric form, in order to explore the most suitable
evolutionary history which reproduces the present observational results.
Furthermore the constraint from Cosmic Infrared Background (CIRB) 
should be considered as another observational constraint to the 
models of evolutionary history.

\subsubsection{CIRB}

First, we searched the evolutionary pattern which satisfy 
the constraint required from CIRB and 
we found three patterns (Evolution $1 - 3$) shown in Fig. 3.
These three evolutionary patterns satisfy the constraint 
from CIRB (Fig. 4).
In order to satisfy the high background intensity 
at $\sim 150\; \mu$m, 
the high evolutionary factor at $z \sim 0.8$ is a mandatory.
We note that too large 
evolutionary factor at high $z$ would produce the excess 
around $1000\; \mu$m.

We obtain IR $60\; \mu$m luminosity density
along with redshift (Fig. 5) from Fig. 3.
Figure 5 show the rapid evolution in $\rho_{\rm L} (60\; \mu{\rm m})$.
The increase is well described by $(1 + z)^5$~!
We need such a sudden rise of $\rho_{\rm L} (60\; \mu{\rm m})$ 
to reproduce the very high CIRB intensity at 
$\sim 150\; \mu$m mentioned before. 
The peak of the IR luminosity density is 
located at $z \sim 1$.

\subsubsection{number count}

Recently, new observational results of the galaxy number counts at the 
mid-infrared and far-infrared (mainly by {\sl ISO}), and submillimeter 
(SCUBA and others).
We are able to compile these data as well as previously obtained 
{\sl IRAS} data and radio data.
It is obviously important to calculate multiband number count predictions 
and compare the multiband observations, because the response of the results 
to the galaxy evolutionary form varies with different wavelengths.
In principle, the galaxy number count is an integrated value along with 
redshift $z$, and the information of the redshift is not available.
But the redshift degeneracy can be solved to some extent, 
by treating the multiband
observational results at the same time.

We check whether the three evolutionary patterns found 
in the previous section also satisfy the constraints from 
observations of number counts.
We compare our number counts with observations in Fig. 6
(infrared), and Fig. 7 (sub-mm -- radio).
Every pattern satisfys the constraints.
When we especially focus on the sub-mm number counts, 
the evolution 3 is the most desirable.

\subsection{Infrared Imaging Surveyor ({\sl IRIS})}

Infrared Imaging Surveyor ({\sl IRIS}) is a satellite 
which will be launched in 2003, by the M-V rocket of the 
{\sl ISAS} (the Institute of Space and Astronautical Science 
in Japan).
One of the main purposes of the {\sl IRIS} mission is 
an all-sky survey at far-infrared (FIR) with a flux limit 
much deeper than that of {\sl IRAS}.
Detailed information of {\sl IRIS} is available 
at http://koala.astro.isas.ac.jp/Astro-F/index-e.html.

In order to examine the performance of the survey,
we estimated the FIR galaxy counts in two narrow bands 
(i.e. N60 and N170) based on models described 
in the previous section.
We found that a large number of galaxies 
( $\sim 10^7$ in the whole sky) will be 
detected in this survey (Fig. 8).

\begin{acknowledgements}

We wish to thank Dr. Hiroshi Shibai, Dr. Izumi Murakami and 
Dr. Hideo Matsuhara for helpful discussions.
Dr. Kimiaki Kawara also deserves our thanks for providing us their 
number count in the Lockman Hole for reference.
TTI is grateful to Dr. Hiroki Kurokawa for continuous encouragement.
TTT, HH, and KY acknowledge the Research Fellowships
of the Japan Society for the Promotion of Science for Young
Scientists.
We are grateful to all the participants who gave us 
useful suggestions and comments at the Symposium.

\end{acknowledgements}

\end{document}